\documentclass[final]{aipproc}

\layoutstyle{8x11double}

\begin{document}

\title{Grain size dependence of barchan dune dynamics}


\author{C. Groh}{
  address={Experimentalphysik V, Universit\"at Bayreuth, Germany}
}

\author{N. Aksel}{
  address={Technische Mechanik und Str\"omungsmechanik, Universit\"at Bayreuth, Germany}
}

\author{I. Rehberg}{
  address={Experimentalphysik V, Universit\"at Bayreuth, Germany}
}

\author{C. Kruelle}{
  address={Experimentalphysik V, Universit\"at Bayreuth, Germany}
  ,altaddress={Maschinenbau und Mechatronik, Hochschule Karlsruhe - Technik und Wirtschaft, Germany}
}

\begin{abstract}
 The dependence of the barchan dune dynamics on the size of the grains involved is investigated experimentally. Downsized barchan dune slices are observed in a narrow water flow tube. The relaxation time from an initial symmetric triangular heap towards an asymmetric shape attractor increases with dune mass and decreases with grain size. The dune velocity increases with grain size. In contrast, the velocity scaling and the shape of the barchan dune is independent of the size of the grains.
\end{abstract}

\maketitle


Our earth finds itself faced with the unpreventable climate change \cite{IPCC2007}. The global warming makes the ice of the polar region melt and the deserts expand. To prevent further desertification, it is necessary to understand the dynamics of sand migration. The interaction between the moving sand and the driving air flow is highly complicated. Therefore it is desirable to investigate interactions of simple model fluid-grain systems.

One of the fundamental dune systems is a barchan. Barchan dunes are crescent-shaped. They can be found as isolated dunes in unidirectional wind fields on bedrock, where sand is sparse \cite{sauermann2000}. These boundary conditions make their dynamics comprehensible for modeling, particularly if the three-dimensional dune is cut in two-dimensional slices along the wind direction. Without losing the basic dynamics this simplification is done with so-called 'minimal models', which deal with two-dimensional barchan dunes \cite{andreotti2002b,kroy2002a,kroy2002b,parteli2006}.
According to these models, the velocity scales reciprocal with the square root of the mass of the barchan dune \cite{andreotti2002b} and reciprocal with the length of the windward side of the dune \cite{kroy2002a,kroy2002b}. Moreover, they predict the existence of a barchan dune attractor \cite{kroy2005,fischer2008}, which has recently been demonstrated experimentally \cite{groh2008b}.

To test these minimal models it is necessary to study barchan dunes in well-controlled experiments in the laboratory. For a direct comparison with the two-dimensional models, it is useful to perform corresponding two-dimensional experiments. In former experiments we were able to verify the prediction of the velocity scaling laws of the minimal-models \cite{groh2008a}. In those experiments the grain size of the dune's granulate has been kept fixed. In this paper, we address the question whether these results are still generally valid if the grain size is changed.

\begin{figure}
\includegraphics[width = 0.9\linewidth]{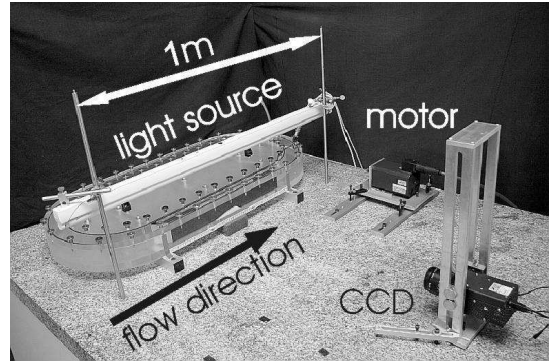}
\caption{Photograph of the experimental setup.}
\label{fig1}
\end{figure}

A photo of our experimental setup is shown in Fig.\,\ref{fig1}.
The main part consists of a closed flow tube machined from perspex which is filled with
distilled water. The height of the channel amounts to 60\,mm, and its width is 50\,mm. The length of the
straight section is 600\,mm and the curves have an outer diameter of 500\,mm.
The water flow is driven by a propeller with a diameter of 45\,mm, which is
installed in the curve following the section of measurements. The flow direction is counter-clockwise on top view.

We limit the section of measurements to approximate two-dimensional conditions, comparable to a Hele-Shaw cell \cite{heleshaw1898}. For that purpose we use a black plastic insert, which constricts the channel to a width $w = 6$\,mm. The aspect ratio of width to height amounts to 10.

For the calculation of the Reynolds number \textrm{Re} of the 6\,mm wide
gap we measure the flow velocity $v_{\textrm{\footnotesize{flow}}}$ with an ultrasonic Doppler velocimeter
(\emph{Signal Processing SA}) and the kinematic viscosity $\nu = 1$\,mm$^{2}$/s of
water at a temperature of $21.7 \pm 0.6$\symbol{23}C. The Reynolds number is kept constant for all measurements at $\textrm{Re}= v_{\textrm{\footnotesize{flow}}}w / \nu = 27000$.

After the flume is filled with distilled water, glass beads are poured into the channel with a funnel at a fixed position upstream. The glass beads have a density of $\rho = 2.5$\,g/cm$^{3}$ and spherical shape. For the following experiments different grain size mixtures are used with diameters $d$
ranging between 400-425\,$\mu$m, 450-500\,$\mu$m, 560-600\,$\mu$m, 710-750\,$\mu$m, and 800-850\,$\mu$m.
The selected range allows the comparison between the different grain sizes at a fixed Reynolds number.
At larger grain sizes $v_{\textrm{\footnotesize{flow}}}$ has to be increased to trigger grain motion, and vice versa.
The diameters are separated by multiple sieving. For the corresponding mean diameters $\bar{d}$ we obtain 412\,$\mu$m, 475\,$\mu$m, 580\,$\mu$m, 725\,$\mu$m, and 825\,$\mu$m, respectively. The different sizes of the barchan dunes are selected
by weighing different amounts of glass beads. The selected masses
$m$ amount to 2.17\,g, 3.25\,g, 6.5\,g, 9.75\,g, 13\,g, 16.25\,g, and 19.5\,g,
with an error of $\pm$0.005\,g.

\begin{figure}
\includegraphics[width = 0.9\linewidth]{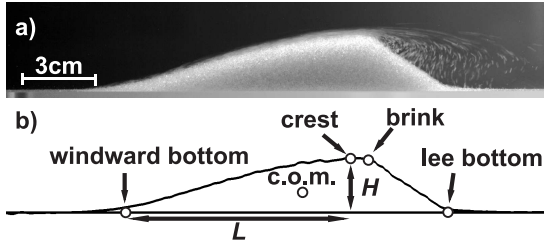}
\caption{Panel a) shows a side view snapshot of a fully developed barchan dune with mass $m$ =
13\,g. The glass beads appear white in front of a dark background. In panel b) the border of the dune is drawn as a black line. Five characteristic points of a
barchan are shown as dots, and the length $L$ and the height $H$ are
plotted as double arrows.}
\label{fig2}
\end{figure}

A charge-coupled device (CCD) camera (\emph{Lumenera} Lw11059) with a horizontal
resolution of $4008\times2672$ pixels and a maximum frame rate of 5\,fps is used for the measurements.
The camera is placed in front of the straight part of the channel and records side views of the glass bead heaps and the developing barchan dunes, respectively. The lighting is done with a luminescent tube above the gap of the section of measurements to illuminate the upper surface.

From pictures like Fig.\,\ref{fig2}a)
the height profile of the barchan is extracted by finding the lowest gradient from dark
to bright. In Fig.\,\ref{fig2}b) the border of the dune is drawn as a black line. We
also indicate five dots which represent the characteristic positions of a
barchan: windward-bottom, crest, brink, lee-bottom and the center of
mass (c.o.m.). The centroid of the two-dimensional
border line indicates the center of mass of the barchan and serves to determine the mean dune velocity $\bar{v}$.
The other four points are determined by using the first and
second derivative of the smoothed height profile, obtained by utilizing
a Gaussian filter. The crest is located at the highest point of the dune.
The brink lies at the maximum of
the second derivative. The windward- and lee-bottom lie at the
minima of the second derivative.

The experiments always start with an elevated triangular heap.
During the time of measurements the heap becomes lower and longer until its shape reaches its steady-state form.
It has been shown that the migration velocity of the dune takes a constant value instantaneously. Mass conservation is maintained all the time \cite{groh2008a}.
Moreover, the relaxation time towards the attractor increases with mass and decreases with the flow velocity of the driving water flow. The final barchan dune attractor does not have the same shape for different flow
velocities and different barchan dune sizes, that means that there is no scale invariance \cite{andreotti2002b}.

The final barchan dune attractor is reached when the ratio $H/L$ becomes constant.
The height $H$ and the length $L$ are defined as shown in Fig.\,\ref{fig2}b).
To get a quantitative statement about the steady state, we fit the
empirical function
\begin{equation} \label{fit1}
\frac{H}{L} = \frac{H_{\infty}}{L_{\infty}} + a e^{-\frac{t}{\tau}}
\end{equation}
to the temporal evolution of $H/L$ \cite{groh2008a}.
The attractor is represented by the offset $H_{\infty}/L_{\infty}$ of
Eq.\,(\ref{fit1}). The aspect ratio $a$ describes the deviation
of the initial aspect ratio from its steady state value $H_{\infty}/L_{\infty}$.
The parameter $\tau$ is the relaxation time for a triangular heap to achieve the
stable shape of a steady-state barchan dune.

\begin{figure}
\includegraphics[width = 0.9\linewidth]{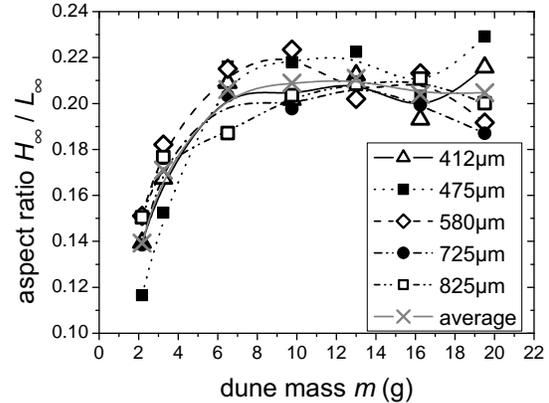}
\caption{Final aspect ratio
$H_{\infty}/L_{\infty}$ for different
barchan sizes and different mean grain diameters $\bar{d}$. The crosses denote the averaged final aspect ratio for each barchan mass $m$. The solid line is a guide to the eye.}
\label{fig3}
\end{figure}

The dependence of the final aspect ratio $H_{\infty}/L_{\infty}$
on the mean grain diameter $\bar{d}$ as a function of the size of the barchan dunes is shown in Fig.\,\ref{fig3}.
The aspect ratio $H_{\infty}/L_{\infty}$ depends on the dune size. To get a better measure we averaged the final aspect ratios for each barchan size.
The resulting mean aspect ratios are plotted as crosses in Fig.\,\ref{fig3}.
For each grain mixture the deviation of $H_{\infty}/L_{\infty}$ from the corresponding mean value seems to be random.
So we suppose that there might be no coupling of the barchan shape, represented by $H_{\infty}/L_{\infty}$, and the grain size.

\begin{figure}
\includegraphics[width = 0.9\linewidth]{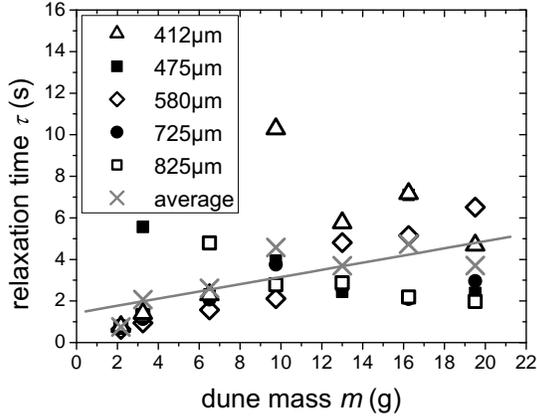}
\caption{Relaxation time $\tau$ for different
barchan sizes and different mean grain diameters $\bar{d}$. The crosses denote the averaged relaxation time $\bar{\tau}_{m}$ for each barchan size. The solid line is a linear fit.}
\label{fig4}
\end{figure}

In Figure\,\ref{fig4} the relaxation time $\tau$ is plotted as a function of the barchan size. To get a qualitative statement we average $\tau$ for each dune size. For the seven different masses explored the resulting $\bar{\tau}_{m}$ is shown as crosses in Fig.\,\ref{fig4}, and a linear fit is plotted to guide the eye. We suggest that for all different grain size mixtures the relaxation time increases with the mass of the dune.

\begin{figure}
\includegraphics[width = 0.9\linewidth]{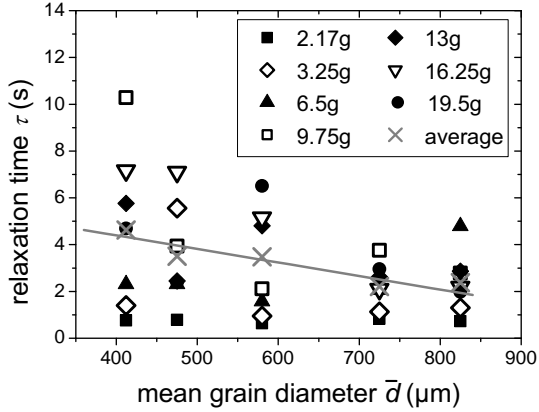}
\caption{Relaxation time $\tau$ as a function of the mean grain diameter $\bar{d}$ for different barchan dune masses $m$. The crosses denote the averaged relaxation time $\bar{\tau}_{\bar{d}}$ for each grain size. The solid line is a linear fit.}
\label{fig5}
\end{figure}

To check the dependence of the relaxation time on the mean grain diameter we average $\tau$ for each grain size mixture. The resulting $\bar{\tau}_{\bar{d}}$ and the original data is plotted as a function of $\bar{d}$ in Fig.\,\ref{fig5}. The mean relaxation time decreases with increasing mean grain diameter.

\begin{figure}
\includegraphics[width = 0.9\linewidth]{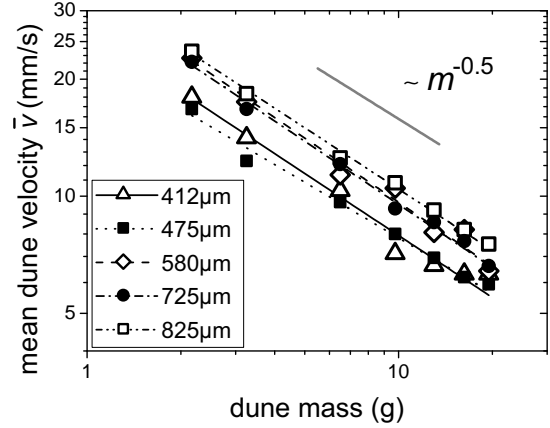}
\caption{Relation between mean barchan velocity $\bar{v}$ and
the mass of the barchan dune $m$ for different mean grain diameters $\bar{d}$.
The lines are fits to $\bar{v} \sim m^{-\alpha}$.}
\label{fig6}
\end{figure}

In Fig.\,\ref{fig6} the
mean dune velocity is plotted versus the mass of the barchan dune. We fit our experimental data to the function $\bar{v} \sim m^{-\alpha}$. The resulting values for $\alpha$ are shown in Table\,\ref{tab}. For
the mean value we can assume $\alpha \approx 0.5$ for all observed grain size mixtures. This corresponds to the theoretical prediction \cite{andreotti2002b} and our former experimental result for a fixed grain size of $d =$ 560-600\,$\mu$m and different flow velocities \cite{groh2008a}.

\begin{figure}
\includegraphics[width = 0.9\linewidth]{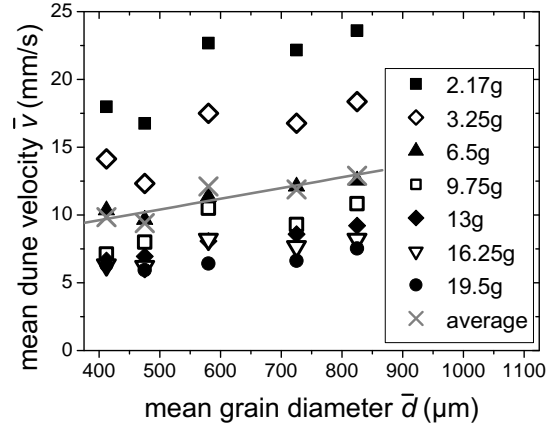}
\caption{Mean barchan velocity $\bar{v}$ as a function of the mean grain diameter $\bar{d}$ for different barchan dune masses $m$. The crosses denote the averaged mean barchan velocity $\bar{v}^{*}$ for each grain size. The solid line is a linear fit.}
\label{fig7}
\end{figure}

For the investigation of the dependence of the barchan velocity on the mean grain diameter we average $\bar{v}$ for each grain size mixture. The resulting $\bar{v}^{*}$ and the original data is plotted as a function of $\bar{d}$ in Fig.\,\ref{fig7}. The mean barchan velocity increases with the mean grain diameter. This results from the fluidized surface layer whose thickness increases with the particle diameter $d$ \cite{bagnold1941}. The larger the fluidized layer, the greater the amount of mass, which is set in motion and shifts the center-of-mass. This also decreases the relaxation time as shown in Fig.\,\ref{fig5}.

\begin{figure}
\includegraphics[width = 0.9\linewidth]{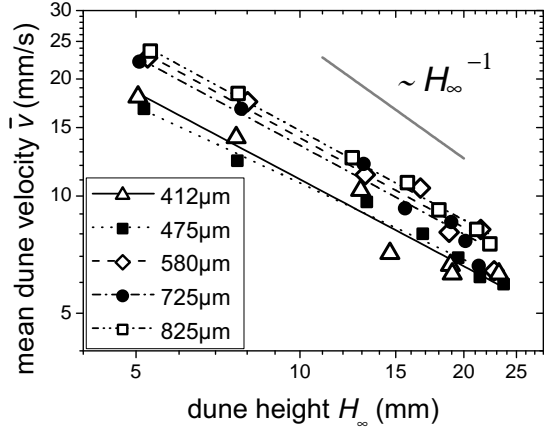}
\caption{Relation between mean barchan velocity $\bar{v}$ and
the steady-state height of the barchan dune $H_{\infty}$ for different mean grain diameters $\bar{d}$. The lines are fits to $\bar{v} \sim H_{\infty}^{-\beta}$.}
\label{fig8}
\end{figure}

\begin{figure}
\includegraphics[width = 0.9\linewidth]{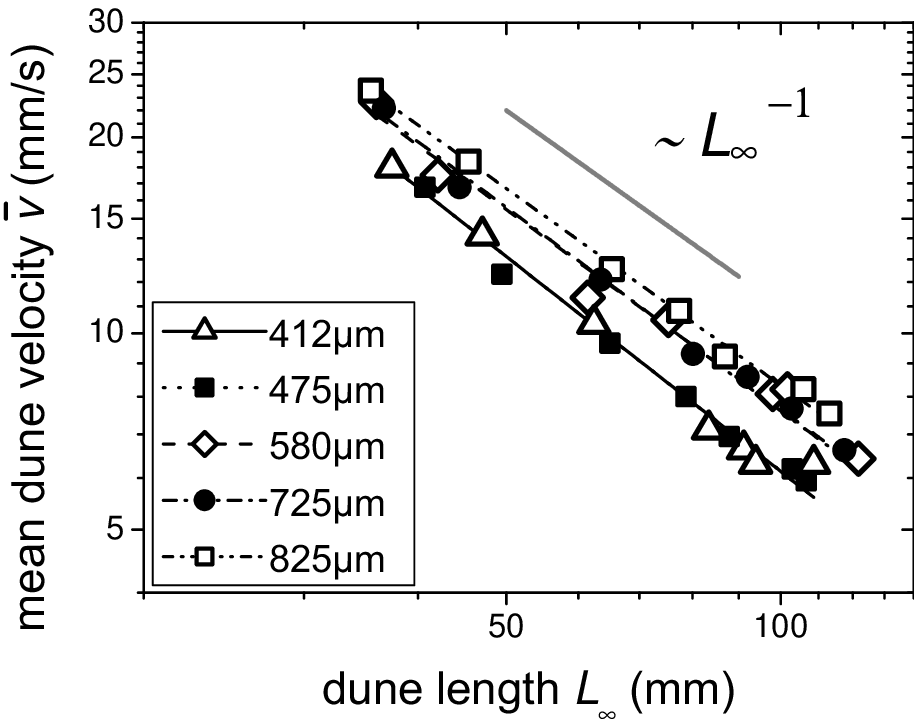}
\caption{Relation between mean barchan velocity $\bar{v}$ and
the steady-state length of the barchan dune $L_{\infty}$ for different mean grain diameters $\bar{d}$.
The lines are fits to $\bar{v} \sim L_{\infty}^{-\gamma}$.}
\label{fig9}
\end{figure}

\begin{table}
\caption{\label{tab}Fit parameters obtained from the experimental data shown in
Fig.\,\ref{fig6}, Fig.\,\ref{fig8}, and Fig.\,\ref{fig9}.}
\begin{tabular}{lrrr}
\tablehead{1}{r}{b}{$\bar{d}$ ($\mu$m)} &
\tablehead{1}{r}{b}{$\alpha$} &
\tablehead{1}{r}{b}{$\beta$} &
\tablehead{1}{r}{b}{$\gamma$} \\
\hline
$825$ & $0.52\pm 0.01$ & $0.76 \pm 0.02$ & $1.00 \pm 0.02$\\
$725$ & $0.54 \pm 0.02$ & $0.76 \pm 0.04$ & $1.04 \pm 0.04$\\
$580$ & $0.55 \pm 0.03$ & $0.77 \pm 0.04$ & $1.02 \pm 0.07$\\
$475$ & $0.47 \pm 0.03$ & $0.65 \pm 0.03$ & $1.08 \pm 0.05$\\
$412$ & $0.53 \pm 0.03$ & $0.74 \pm 0.06$ & $1.10 \pm 0.04$\\
\hline
average & $0.52 \pm 0.02$ & $0.74 \pm 0.05$ & $1.05 \pm 0.04$\\
\end{tabular}
\end{table}

Figure \ref{fig8} and Figure \ref{fig9} show the relations between
dune velocity and $H_{\infty}$, respectively $L_{\infty}$, for
different grain size mixtures. To get a quantitative relation we use the
functions $\bar{v} \sim H_{\infty}^{-\beta}$ and $\bar{v} \sim L_{\infty}^{-\gamma}$ as fits to the data points.

Table\,\ref{tab} shows the results for the exponents $\beta$ and
$\gamma$. From these results we can assume that $\beta = 0.74$
and $\gamma = 1.05$ within the errors of the measurement. This result corresponds to our former experiments with one fixed grain size and different flow velocities, on which the scaling laws $\bar{v} \sim H_{\infty}^{-0.8}$ and $\bar{v} \sim L_{\infty}^{-1}$ are based \cite{groh2008a}.

To conclude, we investigated the dependence of the barchan dune shape attractor on its grain size mixture.
We show experimentally that the relaxation time from a triangular heap towards the shape attractor decreases with the mean diameter of the grains and increases with the overall size of the dune. The dune velocity increases with the grain size. Moreover, the scaling laws for the velocity are independent of the grain size, as well as the shape of the attractor itself.

The next step for understanding the barchan dune dynamics might be the investigation of the surrounding velocity field. Particularly the dependence of the surface roughness on the grain size and the coupling of surface and velocity field must be clarified.

We are grateful for support from Deutsche Forschungsgemeinschaft through Ak13/12-1 and Kr1877/3-1 (Forschergruppe 608 'Nichtlineare Dynamik komplexer Kontinua').

\bibliographystyle{aipproc}   

\bibliography{references}

\begin{thebibliography}{12}
\expandafter\ifx\csname natexlab\endcsname\relax\def\natexlab#1{#1}\fi
\providecommand{\enquote}[1]{``#1''}
\expandafter\ifx\csname url\endcsname\relax
  \def\url#1{\texttt{#1}}\fi
\expandafter\ifx\csname urlprefix\endcsname\relax\def\urlprefix{URL }\fi
\providecommand{\eprint}[2][]{\url{#2}}

\bibitem[IPCC(2007)]{IPCC2007}
IPCC, \emph{Climate Change 2007: The Physical Science Basis}, Cambridge
  University Press, Cambridge, 2007.

\bibitem[Sauermann et~al.(2000)]{sauermann2000}
G.~Sauermann, P.~Rognon, A.~Poliakov, and H.~J. Herrmann, \emph{Geomorphology}
  \textbf{36}, 47--62 (2000).

\bibitem[Andreotti et~al.(2002)]{andreotti2002b}
B.~Andreotti, P.~Claudin, and S.~Douady, \emph{Eur. Phys. J. B} \textbf{28},
  341--352 (2002).

\bibitem[Kroy et~al.(2002{\natexlab{a}})]{kroy2002a}
K.~Kroy, G.~Sauermann, and H.~J. Herrmann, \emph{Phys. Rev. Lett.} \textbf{88},
  054301 (2002{\natexlab{a}}).

\bibitem[Kroy et~al.(2002{\natexlab{b}})]{kroy2002b}
K.~Kroy, G.~Sauermann, and H.~J. Herrmann, \emph{Phys. Rev. E} \textbf{66},
  031302 (2002{\natexlab{b}}).

\bibitem[Parteli et~al.(2006)]{parteli2006}
E.~J.~R. Parteli, V.~Schw\"{a}mmle, H.~J. Herrmann, L.~H.~U. Monteiro, and
  L.~P. Maia, \emph{Geomorphology} \textbf{81}, 29--42 (2006).

\bibitem[Kroy et~al.(2005)]{kroy2005}
K.~Kroy, S.~Fischer, and B.~Obermayer, \emph{J. Phys.: Condens. Matter}
  \textbf{17}, S1229 (2005).

\bibitem[Fischer et~al.(2008)]{fischer2008}
S.~Fischer, M.~E. Cates, and K.~Kroy, \emph{Phys. Rev. E} \textbf{77}, 031302
  (2008).

\bibitem[Groh et~al.(2008{\natexlab{a}})]{groh2008b}
C.~Groh, I.~Rehberg, and C.~A. Kruelle, \emph{New Journal of Physics}
  submitted (2008{\natexlab{a}}).

\bibitem[Groh et~al.(2008{\natexlab{b}})]{groh2008a}
C.~Groh, A.~Wierschem, N.~Aksel, I.~Rehberg, and C.~A. Kruelle, \emph{Phys.
  Rev. E} \textbf{78}, 021304 (2008{\natexlab{b}}).

\bibitem[Hele-Shaw(1898)]{heleshaw1898}
H.~Hele-Shaw, \emph{Transactions of the Institute of Naval Architects}
  \textbf{40}, 21--46 (1898).

\bibitem[Bagnold(1941)]{bagnold1941}
R.~A. Bagnold, \emph{The Physics of Blown Sand and Desert Dunes}, Chapman and
  Hall, London, 1941.

\end{thebibliography}

\IfFileExists{\jobname.bbl}{}
 {\typeout{}
  \typeout{******************************************}
  \typeout{** Please run "bibtex \jobname" to optain}
  \typeout{** the bibliography and then re-run LaTeX}
  \typeout{** twice to fix the references!}
  \typeout{******************************************}
  \typeout{}
 }

\end{document}